# Manipulation of Diatomic Molecules with "Oriented External Electric Fields": Linear Correlations in Atomic Properties Lead to Non-Linear Molecular Responses

Shahin Sowlati-Hashjin,[a,b,c,d] Mikko Karttunen,[a,b,e] Chérif F. Matta[c,d,f]*

[a] *Department of Chemistry, The University of Western Ontario, 1151 Richmond Street, London, Ontario N6A 5B7, Canada.* [b] *The Centre of Advanced Materials and Biomaterials Research, The University of Western Ontario, 1151 Richmond Street, London, Ontario N6A 5B7, Canada.* [c] *Department of Chemistry, Saint Mary's University, Halifax, Nova Scotia, B3H 3C3, Canada.* [d] *Department of Chemistry and Physics, Mount Saint Vincent University, Halifax, Nova Scotia, B3M 2J6, Canada.* [e] *Department of Applied Mathematics, The University of Western Ontario, 1151 Richmond Street, London, Ontario, N6A 5B7, Canada.* [f] *Department of Chemistry, Dalhousie University, Halifax, Nova Scotia, B3H,4J3, Canada.*

*\* Tel.: +1-(902)-457-6142, Fax: +1-(902)-457-6134, E-mail: cherif.matta@msvu.ca.*

**TOC GRAPHIC**

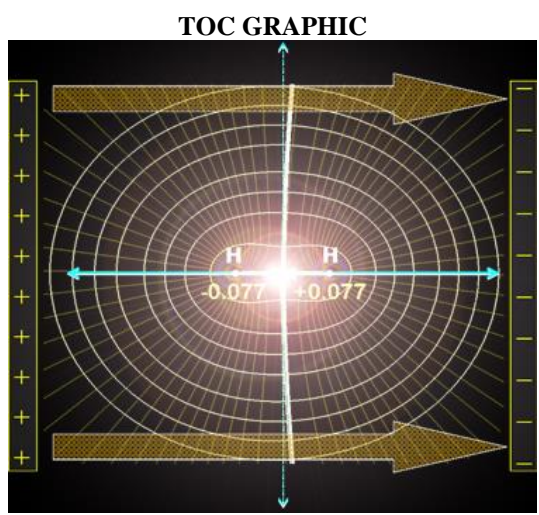

**Abstract**

"Oriented external electric fields (OEEFs)" have been shown to have great potential in being able to provide unprecedented control of chemical reactions, catalysis and selectivity with applications ranging from $H_2$ storage to molecular machines. We report a theoretical study of the atomic origins of molecular changes due to OEEFs; understanding the characteristics of OEEF-induced couplings between atomic and molecular properties is an important step toward comprehensive understanding of the effects of strong external fields on molecular structure, stability, and reactivity. We focus on the atomic and molecular (bond) properties of a set of homo- ($H_2$, $N_2$, $O_2$, $F_2$, and $Cl_2$) and hetero-diatomic (HF, HCl, CO, and NO) molecules under intense external electric fields in the context of quantum theory of atoms in molecules (QTAIM). It is shown that *atomic properties* (atomic charges and energies, and localization index) *correlate linearly* with the field strengths, *but molecular properties* (bond length, electron density at bond critical point, bond length, and electron delocalization index) *exhibit non-linear responses* to the imposed fields. In particular, the changes in the electron density distribution alter the shapes and locations of the zero-flux surfaces, atomic volumes, atomic electron population, and localization/delocalization indices. At the molecular level, the topography and topology of the molecular electrostatic potential undergo dramatic changes. The external fields also perturb the covalent-polar-ionic characteristic of the studied chemical bonds, hallmarking the impact of electric fields on the stability and reactivity of chemical compounds. The findings are well-rationalized within the framework of the quantum theory of atoms in molecules and form a coherent conceptual understanding of these effects in prototypical molecules such as diatomics.

1. **Introduction**

Intense electric fields are ubiquitous in the microenvironments of molecules from enzyme active sites[1] to precise control of enantioselectivity,[2] nano-circuitry,[3] and scanning tunneling microscopy.[4] For example, fields of magnitudes in the range of $10^8$ to $10^9$ V.m$^{-1}$ are commonly found in the active sites of enzymes and they are instrumental to their catalytic functions.[5-13] Changes in these fields upon mutating an amino acid residue can be monitored through vibrational Stark shifts of small host molecules such as carbon monoxide attached to the iron of the porphyrin ring in myoglobin[14] or nitrile-containing substrates in the active site of human aldose reductase enzyme (hALR2),[15-16] and thus accurate modeling of electric fields in the active site cavity of an enzyme is a requirement for prediction of mutation-induced Stark shift.[17-19] In addition, oriented external electric fields offer the potential to control chemical reactions at single molecule level[11, 20] and their application for selective and controlled catalysis has been demonstrated.[12, 20] Recent reviews are regarding the effects of electric (and other fields) are provided, for example by Shaik et al.,[13] Fried and Boxer,[5] and Sato.[21]

It is well-established that strong external electric fields, $10^9$ V.m$^{-1}$ and higher, can significantly alter the topography of potential energy surfaces (PES),[5-13] molecular charge density distributions[22-23] and equilibrium bond lengths as well as vibrational frequencies, see, e.g., Refs. [24-25] and references therein. These changes are manifested in diverse manners including enhancements in probabilities of particular reaction paths,[7] changes in the rates of hydrogen transfer reactions,[26-28] changes in the energetics of $H_2$ formation at the active sites of [FeFe] hydrogenases,[29] controlling isomerization and molecular switches[3, 30-32] and in improving the performance of switches by weakening the hydrogen bonds,[33] and some uncommon reactions.[34] Such electric fields can also accelerate photosynthetic reactions[35-37] and decrease the enzymatic activity of cytochrome *c* oxidase.[38] Regarding other applications, being able to control field-induced modulations of the PES has been proposed as a new approach to enhance the loading and unloading of $H_2$ molecules on polarizable nano-sheets with the goal of developing field enhanced hydrogen storage technology.[39-40] This also underlines the need to understand the effects of OEEFs on diatomic molecules.

We study the effects of OEEFs on atomic and molecular properties, and their field induced couplings using nine diatomic molecules: Five homo-nuclear diatomics with a wide range of polarizabilities ($H_2$, $N_2$, $O_2$, $F_2$, and $Cl_2$) and four hetero-nuclear diatomic molecules of varying permanent dipole moments and polarizabilities (CO, HF, NO, and HCl).[25] The aim is to decipher the atomic origins of molecular changes caused by OEEFs to understand the atomic contributions to the observed changes in molecular structure and properties. It is shown that electric fields affect the *nature* of the chemical bonds through their effects on the polarities of the molecules. Fundamental chemical properties such as nucleophilicity and electrophilicity of the molecule may be manipulated through altering the polarity of chemical bonds, providing means to control specific chemical reactions.

We use the field-perturbed Morse potential previously proposed[25] and Bader's Quantum Theory of Atoms in Molecules (QTAIM),[41-44] as a framework of this study. This approach allows for prediction of the changes caused by external electric fields on the equilibrium bond lengths and vibrational Stark shifts of diatomic molecules.[25] The field-perturbed Morse potential approach has been previously validated and verified against direct brute force calculations of vibrational frequencies and bond lengths.[25] The task now is to relate the field-induced changes in molecular and bond properties to changes in the properties of the atoms themselves, that is, to the field-induced changes in atomic properties.

## 2. Computational Details

Below, we provide a brief recap of the computational approach. Full details have been described elsewhere.[25] Molecules are modelled in their electronic ground states which are generally $^1\Sigma_g^+$ and $^1\Sigma^+$, except for NO ($^2\Pi_r$) and for O$_2$ ($^3\Sigma_g^-$). As per standard spectroscopic notation, $\Sigma$ and $\Pi$ stand for the total angular momentum of 0 and 1, respectively, and the letter $g$ represents total parity (gerade = even), and ± sign characterizes the change of wavefunction under upon reflection in a plane containing the internuclear axis where (+) denotes a symmetric wavefunction and (−) means an antisymmetric wavefunction for closed-shell homo-nuclear and hetero-nuclear diatomics, respectively. The superscripts 2 and 3 designate one (↑) and two (↑ ↑) unpaired electrons in two degenerate orbitals, respectively. The geometries and wave functions for these molecules, taken from our previous study,[25] are obtained in the (truncated) quadratic configuration interaction approximation with single and double excitations (QCISD)[45] in conjunction with the Pople-type basis set 6-311++G(3$df$,2$pd$); and whereby the unrestricted formalism is used for open-shell molecules. Electronic structure calculations and geometry optimizations were performed using the Gaussian 09 software.[46]

Electric fields were imposed on the molecules in both parallel and antiparallel directions with respect to their inter-nuclear axes. When applicable, the least electronegative atom was placed at the origin and the second atom in the positive side of the $z$-axis (Figure 1; the same coordinate system and orientations were used as in Ref. [25]). The externally applied fields (all being uniform and time-invariant of a strength of $1.03\times10^{10}$ V.m$^{-1}$ = $2.0\times10^{-2}$ atomic units (a.u.) unless stated otherwise) are either directed positively along the *z*-direction ($\vec{E}_+$) or in the opposite direction ($\vec{E}_-$). Dipole moment vectors are oriented according to the "physicist's convention",[47-48] originating at the negative pole and pointing to the positive pole. As per the usual convention, positive charges are the sources of electric fields and negative charges are sinks (Gaussian 09 uses the opposite convention; see Figure 1).

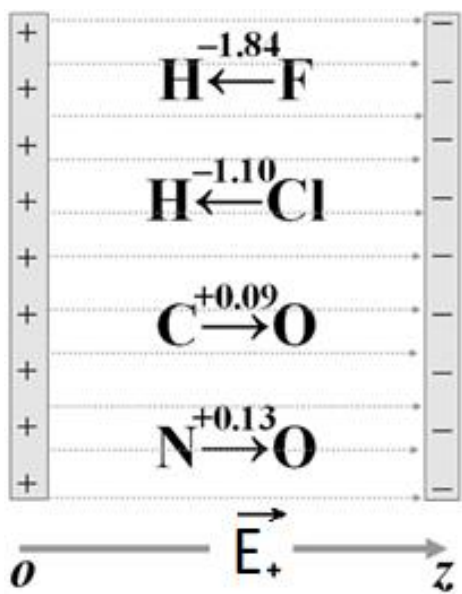

**Figure 1**
Orientation of the permanent dipole moment of hetero-nuclear diatomic molecules with respect to the coordinate system and an external electric field. The direction of the electric field is positive in this figure, pointing to the positive *z*-direction. The arrows between the elemental symbols indicate the direction of the permanent (field-free) molecular dipole moment, in the "physicist convention", with the value of the dipole moment calculated at the QCISD level to two decimals in debyes (the ± sign indicates a parallel/antiparallel orientation with respect to the field, respectively).

The QCISD/6-311++G(3*df*,2*pd*) wave functions were subjected to QTAIM analysis using the AIMAll software[49] and the resulting densities displayed using the associated graphical interface AIMStudio.[49] The sum of the virial atomic energies and the energies calculated directly agree to within an average absolute deviation of 0.01±0.03 kcal.mol$^{-1}$ for the nine studied diatomic molecules with a maximum deviation of 0.09 kcal.mol$^{-1}$ for CO.

**3.     Results and Discussion**

Tables 1 and 2 summarize molecular and atomic properties in the field-free case and in the presence of the strongest field (1.03×10$^{10}$ V.m$^{-1}$ = 2.0×10$^{-2}$ a.u.) – with the field applied in both directions in the case of hetero-diatomic molecules. The details are discussed in the following sections.

*3.1.     Molecular Effects of External Electric Fields*

Table 1 shows that all homo-nuclear diatomic molecules are more stable in the presence of a field. Hetero-nuclear diatomic molecules with strong permanent dipole moment (HF and HCl) become more stable only in the parallel external field while those with weak permanent dipole moments (CO and NO) are stabilized in both orientations. These observations have already been explained elsewhere,[25] but for completeness a brief explanation is provided in the next paragraph.

All homo-nuclear diatomic molecules become increasingly stabilized upon increasing electric fields with a quadratic dependence on the field strength (with Cl$_2$ exhibiting the largest relative stabilization). This is expected from the form of the expression for the field-induced change in the total energy for co-linear fields ($\Delta E$):

$$\Delta E = E_\mathbf{E} - E_0 = \mp \mu_0(r)E - \frac{1}{2}\alpha_{//0}(r)E^2, \tag{1}$$

where $E_0$, $\mu_0$, and $\alpha_{//0}$ are, respectively, the field-free energy, permanent dipole moment, and the field-free parallel polarizability tensor component. The dipole term assumes a negative sign for parallel fields and a positive sign for an antiparallel field. The subscript "0" indicates a field-free quantity, while the subscript "**E**" denotes that a quantity is calculated in the presence of an external field. For a homo-nuclear diatomic molecule the first term in the last R.H.S. of Eq. (1) vanishes leaving only the negative (hence always stabilizing) quadratic term weighted by polarizability. Since $Cl_2$ is the most polarizable molecule in the set, it is the one exhibiting the most pronounced field-effect stabilization.

In the case of hetero-nuclear diatomic molecules, the behavior depends on the relative weights of the dipolar (**μ**) and polarizability (**α**) terms in Eq. (1). Thus, expressing all following dipole moments in debyes and polarizabilities in $Å^3$, a molecule with a relatively large dipole moment and low polarizability, such as HF ($\mu_0$ = -1.8358 calc. (1.8262 exptl.) and $\alpha_{avg.}$ = 0.7123 calc. (0.80 exptl.)[25]), exhibits a quasi-linear dependence on the electric field strength, stabilizing for parallel fields and destabilizing for antiparallel fields. At the other extreme are polarizable molecules with a small permanent dipole (CO ($\mu_0$ = +0.0846 calc. (0.1098 exptl.) and $\alpha_{avg.}$ = 1.9306 calc. (1.95 exptl.)) and NO ($\mu_0$ = +0.1247 calc. (0.1587 exptl.) and $\alpha_{avg.}$ = 1.6441 calc. (1.70 exptl.)[25])). Their behavior approaches that of homo-nuclear diatomic molecules since in these cases the second term in Eq. (1) is dominant (see Ref. [25]). Molecules such as HCl exhibit a turning point since its curve reaches a maximum for antiparallel (positive) fields before the negative polarizability term starts to dominate.[25]

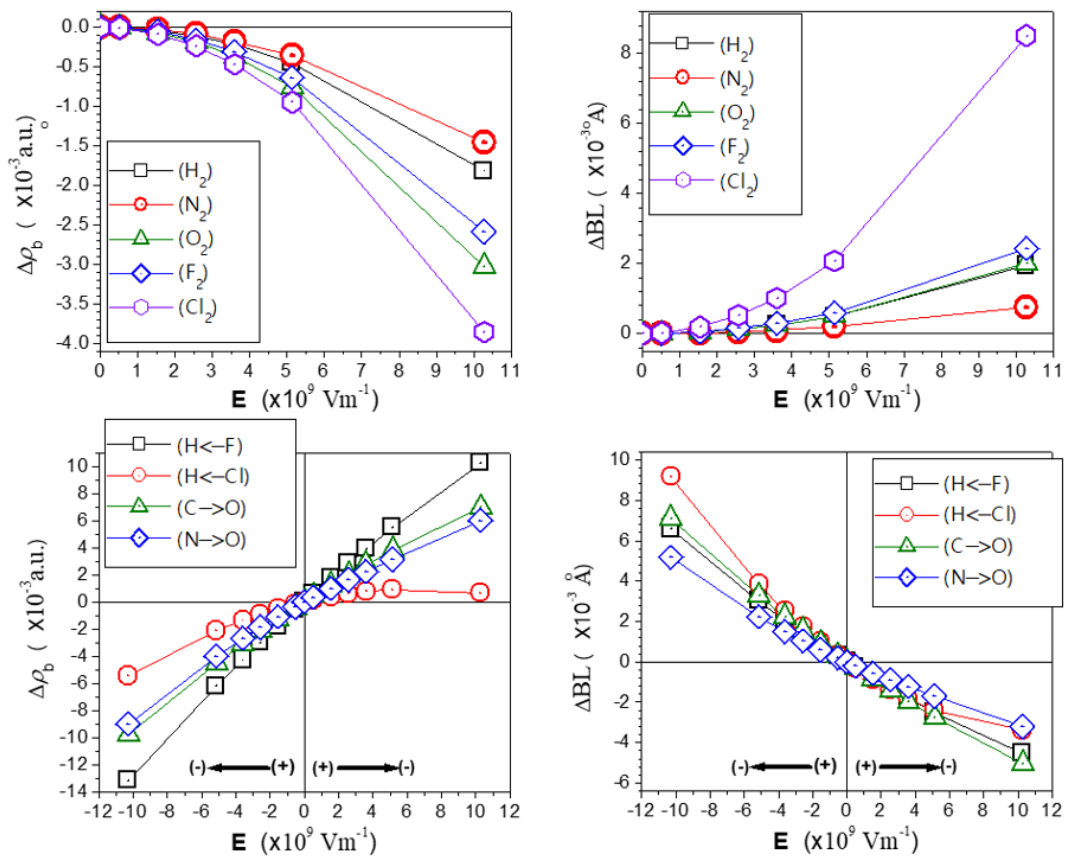

**Figure 2**
The change in the electron density at the bond critical point (BCP), $\Delta\rho_b$ (*left*), and the accompanying change in the bond length $\Delta$BL (*right*), against the strength of the external electric field for homo-nuclear diatomics (*top*) and hetero-nuclear diatomics (*bottom*).

The changes in the electron density at the bond critical point (BCP) induced by the external field, $\Delta\rho_b$, for the nine diatomics are expressed numerically for the strongest field considered in this study in Table 1 and as a function of field strength and direction in Figure 2. The Figure shows that changes in $\rho_b$ are in the opposite direction to the changes in the bond lengths. It is not possible to decouple the direct effect of the field on the electron density at the

**Table 1***     Molecular and bond properties (*P*) of diatomic molecules and their changes (Δ*P*) in electric fields (**E**$_\pm$) of ±1.03×10$^{10}$ V.m$^{-1}$ = ±2.0×10$^{-2}$ a.u.$^{(a)}$

| Property | Field | H$_2$ | N$_2$ | O$_2$ | F$_2$ | Cl$_2$ | H←F$^{(b)}$ | H←Cl$^{(b)}$ | C→O$^{(b)}$ | N→O$^{(b)}$ |
|---|---|---|---|---|---|---|---|---|---|---|
| Δ*E* (kcal/mol) | **E**$_+$ | -0.8 | -1.8 | -1.9 | -1.5 | -5.1 | 8.3 | 3.3 | -2.4 | -2.5 |
| *E* (a.u.) | 0 | **-1.17235** | **-109.35590** | **-150.11162** | **-199.27218** | **-919.39185** | **-100.33424** | **-460.32223** | **-113.14149** | **-129.70083** |
| Δ*E* (kcal/mol) | **E**$_-$ | | | | | | -9.8 | -7.6 | -1.6 | -1.3 |
| ΔBL(Å) | **E**$_+$ | 0.0019 | 0.0008 | 0.0020 | 0.0024 | 0.0085 | -0.0045 | -0.0034 | -0.0051 | -0.0032 |
| BL(Å) | 0 | 0.7426 | 1.0975 | 1.1995 | 1.3938 | 1.9974 | 0.9146 | 1.2736 | 1.1285 | 1.1504 |
| ΔBL(Å) | **E**$_-$ | | | | | | 0.0066 | 0.0092 | 0.0071 | 0.0052 |
| Δ*v* (cm$^{-1}$)$^{(c)}$ | **E**$_+$ | -32.8 | -7.2 | -20.3 | -7.1 | -15.4 | 64.7 | 20.8 | 38.4 | 59.0 |
| *v* (cm$^{-1}$)$^{(c)}$ | 0 | 4405.3 | 2404.5 | 1669.2 | 988.6 | 562.2 | 4207.8 | 3021.0 | 2196.0 | 1996.4 |
| Δ*v* (cm$^{-1}$)$^{(c)}$ | **E**$_-$ | | | | | | -103.4 | -82.0 | -56.0 | -47.1 |
| Δ$\rho_b$ (au) | **E**$_+$ | -0.0018 | -0.0015 | -0.0030 | -0.0026 | -0.0039 | 0.0102 | 0.0007 | 0.0069 | 0.0060 |
| $\rho_b$ (au) | 0 | **0.2685** | **0.6957** | **0.5560** | **0.2957** | **0.1622** | **0.3831** | **0.2575** | **0.5063** | **0.5971** |
| Δ$\rho_b$ (au) | **E**$_-$ | | | | | | -0.0132 | -0.0055 | -0.0098 | -0.0090 |
| Δδ(Ω,Ω') | **E**$_+$ | -0.0050 | -0.0031 | -0.0036 | -0.0032 | -0.0110 | 0.0391 | 0.0533 | 0.0463 | 0.0170 |
| δ(Ω,Ω') | 0 | **0.8443** | **2.3118** | **1.2890** | **0.9025** | **1.1000** | **0.4056** | **0.8603** | **1.4151** | **1.6446** |
| Δδ(Ω,Ω') | **E**$_-$ | | | | | | -0.0362 | -0.0602 | -0.0465 | -0.0208 |

* For each property, the table lists values under the strongest fields in the two opposite directions in the first and third row and the field-free value is in the middle row. For every molecule, the properties investigated as a function of the field include the total energy (*E*), equilibrium bond length (BL), harmonic vibrational frequency (*v*), electron density at the bond critical point ($\rho_b$), and the delocalization index between the two atoms forming the diatomic (δ(Ω,Ω′)). Since for homo-nuclear diatomic molecules the two field directions are equivalent, the fields are assigned a positive sign.

(a) Data calculated at the (U)QCISD/6-311++G(3*df*,2*pd*) level of theory. (b) The arrow between the atomic symbols depicts the direction of the field-free (permanent) dipole moment. Note that this direction may flip sides under a strong external field in the opposite direction [see also footnote (c)]. (c) A negative dipole moment points to the left (−$\mu$ ≡ $\tilde{\boldsymbol{\mu}}$) and one that is positive to the right with respect to the other vectors indicated by arrows in this table [see also footnote (b)].

**Table 2**      Atomic properties ($P$) and their changes ($\Delta P = P_E - P_0$) in electric fields ($\mathbf{E}_\pm$) of $\pm 1.03\times 10^{10}$ V.m$^{-1}$ = $\pm 2.00\times 10^{-2}$ a.u.[a,b]

*Homo-nuclear diatomics*[a]

| | $\delta^+$H(1)–H(2)$^{\delta-}$ | | $\delta^+$N(1)–N(2)$^{\delta-}$ | | $\delta^+$O(1)–O(2)$^{\delta-}$ | | $\delta^+$F(1)–F(2)$^{\delta-}$ | | $\delta^+$Cl(1)–Cl(2)$^{\delta-}$ | |
|---|---|---|---|---|---|---|---|---|---|---|
| E-field | **0** | $\vec{\mathbf{E}}_+$ | **0** | $\vec{\mathbf{E}}_+$ | **0** | $\vec{\mathbf{E}}_+$ | **0** | $\vec{\mathbf{E}}_+$ | **0** | $\vec{\mathbf{E}}_+$ |
| $\Omega_1$ | $P$ | $\Delta P$ | $P$ | $\Delta P$ | $P$ | $\Delta P$ | $P$ | $\Delta P$ | $P$ | $\Delta P$ |
| $E$ | -0.58617 | 17.6 | -54.67795 | 30.4 | -75.05581 | 20.7 | -99.63609 | 16.6 | -459.69593 | 26.4 |
| $q$ | 0.0000 | 0.0765 | 0.0000 | 0.1020 | 0.0000 | 0.0957 | 0.0000 | 0.0584 | 0.0000 | 0.1360 |
| $\Lambda$ | 0.5778 | -0.0741 | 5.8441 | -0.1007 | 7.3555 | -0.0939 | 8.5487 | -0.0568 | 16.4500 | -0.1310 |
| %loc | 57.8 | -3.2 | 83.5 | -0.2 | 91.9 | -0.1 | 95.0 | -0.0 | 96.8 | 0.0 |
| $r_b$ | 0.3713 | -0.0080 | 0.5488 | -0.0011 | 0.5997 | 0.0006 | 0.6969 | -0.0042 | 0.9987 | -0.0191 |
| $\Omega_2$ | | | | | | | | | | |
| $E$ | -0.58617 | -18.4 | -54.67795 | -32.2 | -75.05581 | -22.7 | -99.63609 | -18.1 | -459.69593 | -31.5 |
| $q$ | 0.0000 | -0.0765 | 0.0000 | -0.1020 | 0.0000 | -0.0957 | -0.0000 | -0.0584 | 0.0000 | -0.1360 |
| $\Lambda$ | 0.5778 | 0.0790 | 5.8441 | 0.1038 | 7.3555 | 0.0975 | 8.5487 | 0.0601 | 16.4500 | 0.1420 |
| %loc | 57.8 | 3.2 | 83.5 | 0.3 | 91.9 | 0.1 | 95.0 | 0.1 | 96.8 | 0.1 |
| $r_b$ | 0.3713 | 0.0100 | 0.5488 | 0.0018 | 0.5997 | 0.0014 | 0.6969 | 0.0067 | 0.9987 | 0.0276 |

*Hetero-nuclear diatomics*[b]

| | H(1)←F(2) | | | H(1)←Cl(2) | | | C(1)→O(2) | | | N(1)→O(2) | | |
|---|---|---|---|---|---|---|---|---|---|---|---|---|
| E-field | $\vec{\mathbf{E}}_-$ | **0** | $\vec{\mathbf{E}}_+$ | $\vec{\mathbf{E}}_-$ | **0** | $\vec{\mathbf{E}}_+$ | $\vec{\mathbf{E}}_-$ | **0** | $\vec{\mathbf{E}}_+$ | $\vec{\mathbf{E}}_-$ | **0** | $\vec{\mathbf{E}}_+$ |
| $\Omega_1$ | $\Delta P$ | $P$ | $\Delta P$ | $\Delta P$ | $P$ | $\Delta P$ | $\Delta P$ | $P$ | $\Delta P$ | $\Delta P$ | $P$ | $\Delta P$ |
| $E$ | -22.7 | -100.07039 | 21.9 | -35.9 | -459.81999 | 29.0 | -11.6 | -76.07911 | 10.4 | -18.4 | -75.45784 | 18.5 |
| $q$ | -0.0267 | -0.7488 | 0.0303 | -0.0838 | -0.2639 | 0.0857 | -0.0647 | -1.2196 | 0.0696 | -0.0825 | -0.4389 | 0.0870 |
| $\Lambda$ | 0.0449 | 9.5460 | -0.0498 | 0.1140 | 16.8340 | -0.1130 | 0.0879 | 8.5122 | -0.0929 | 0.0930 | 7.6166 | -0.0955 |
| %loc | 0.2 | 97.9 | -0.2 | 0.2 | 97.5 | -0.2 | 0.3 | 92.3 | -0.3 | 0.2 | 90.3 | -0.2 |
| $r_b$ | 0.0058 | 0.7805 | -0.0119 | 0.0236 | 0.9028 | -0.0187 | 0.0052 | 0.7484 | -0.0036 | 0.0034 | 0.6806 | -0.0033 |
| $\Omega_2$ | | | | | | | | | | | | |
| $E$ | 12.9 | -0.26386 | -13.5 | 28.3 | -0.50223 | -25.7 | 10.0 | -37.06223 | -12.9 | 17.1 | -54.24299 | -21.0 |
| $q$ | 0.0267 | 0.7488 | -0.0303 | 0.0840 | 0.2638 | -0.0855 | 0.0649 | 1.2193 | -0.0692 | 0.0826 | 0.4389 | -0.0870 |
| $\Lambda$ | -0.0087 | 0.0484 | 0.0107 | -0.0539 | 0.3061 | 0.0588 | -0.0418 | 4.0734 | 0.0457 | -0.0721 | 5.7388 | 0.0785 |
| %loc | -1.6 | 19.3 | 1.7 | -2.9 | 41.6 | 2.8 | 0.3 | 85.2 | -0.3 | 0.0 | 87.5 | 0.0 |
| $r_b$ | -0.0046 | 0.1395 | 0.0019 | -0.0144 | 0.3709 | 0.0153 | 0.0019 | 0.3800 | -0.0014 | 0.0018 | 0.4698 | 0.0001 |

(a) For each diatomic, the column labeled with "**0**" lists field-free value $P$ and that labeled with $\mathbf{E}$ ( = 1.03×10$^{10}$ V.m$^{-1}$ in a given direction) lists the change in the property ($\Delta P = P_E - P_0$) under that external electric field. (Data calculated at the (U)QCISD/6-311++G(3df,2pd) level of theory). (b) $E$, $q$, $\Lambda$, %loc, and $r_b$ are the atomic total energy, atomic charge, localization index, localization percentage, and atomic radius, respectively. All data are in atomic units (a.u.) except the changes in the total energies $\Delta E$ which are converted to kcal/mol, and except $r_b$ and $\Delta r_b$ which are given in angstroms (Å). (c) For homo-nuclear diatomics, $\Omega_1^{\delta+}$ refers to the atom closer to the sink of the field lines (the negative infinite plate) and $\Omega_2^{\delta-}$ refers to the atom closer to the source of the field lines (the positive infinite plate). (d) For hetero-nuclear diatomics ($\Omega_1-\Omega_2$), $\Omega_1$ is always the atom on the left of the molecular structure as written and $\Omega_2$ refers the one on the right.

BCP from the indirect effect through elongating the bond. In order to achieve this decoupling, one needs another set of calculations with varying field strengths but using a frozen field-free geometry, which is the subject of a separate study. The homo-nuclear molecule that exhibits the strongest response (deviating the most from linearity) with the field strength is $Cl_2$, while those that show the minimum and the maximum deviations from linearity in their response to the field in the hetero-nuclear set are, respectively, HF and HCl.

## *3.2. Effects of External Electric fields on Electron Density and Molecular Electrostatic Potential (MESP)*

Figure 3 displays a contour plot of the electron density $\rho(r)$ and a representation of the associated gradient vector field $[\nabla\rho(\mathbf{r})]$ lines in a plane containing both nuclei of the $H_2$ molecule in the presence of an external field of $1.03\times10^{10}$ V.m$^{-1}$ parallel to the inter-nuclear axis. This field, while very intense at macroscopic scale, is only ~ 2% of the field strength at the first Bohr orbit of the hydrogen atom (1 a.u. = $e/4\pi\varepsilon_0 a_0^2 \approx 5.14\times10^{11}$ V.m$^{-1}$). Hence, it is expected to slightly perturb the density, but this slight perturbation is amplified energetically through the quadratic term in Eq. (1), especially when a change (rather than the absolute value) in energy is important as in the case of the change in reaction energy barriers.[26-28]

The perturbing effect of this field on the electron density distribution of $H_2$ can be seen from Figure 3. This figure reveals that (*i*) the Bader zero-flux H|H interatomic surface of $H_2$ becomes slightly curved - concave toward the negative pole of the external field and convex toward the positive pole, and that (*ii*) the electron density contours are compressed (less diffuse) facing the negative pole and more diffuse facing the positive plate as can be judged by looking at the respective extensions of contours from the closest nucleus to the plate or from the BCP. To facilitate this comparison, an arrow is drawn from the bond critical point along the bond axis to the outermost contour (the $\rho = 0.001$ a.u. contour) from the side facing the negative plate (the right side). The same arrow displaced to the left can only reach the $\rho = 0.002$ a.u. contour, the 0.001 a.u. contour being visibly more diffuse.

The shift in electron population to the hydrogen basin in the left of Figure 3 (the negatively charged atom) and its more diffuse distribution lead to an increase in its volume from 59.0 a.u. in the field-free $H_2$ to 65.5 a.u., while the atom to the right loses an equal amount of its electron population and shrinks in volume to 54.7 a.u. These drastic changes in atomic volumes in opposite directions cancel to a large extent when summed up to yield the molecular value. Thus, the external field changes the volume of the $H_2$ molecule from 118.0 a.u. to 120.2 a.u. (= 65.5 + 54.7 a.u.), that is, expands the molecular volume by 2.2 a.u. despite the large inter-atomic reorganization of electron populations and volumes of atomic basins. This can be considered as a first example of field-induced (response) "*compensatory transferability*" whereby the system, as a whole, resists change through internally induced changes. Compensatory transferability has a different meaning in the traditional QTAIM literature whereby changes in different atoms in the system tend to cancel one another to sustain transferability of functional groups (see discussions by Bader and Bayles).[50-52]

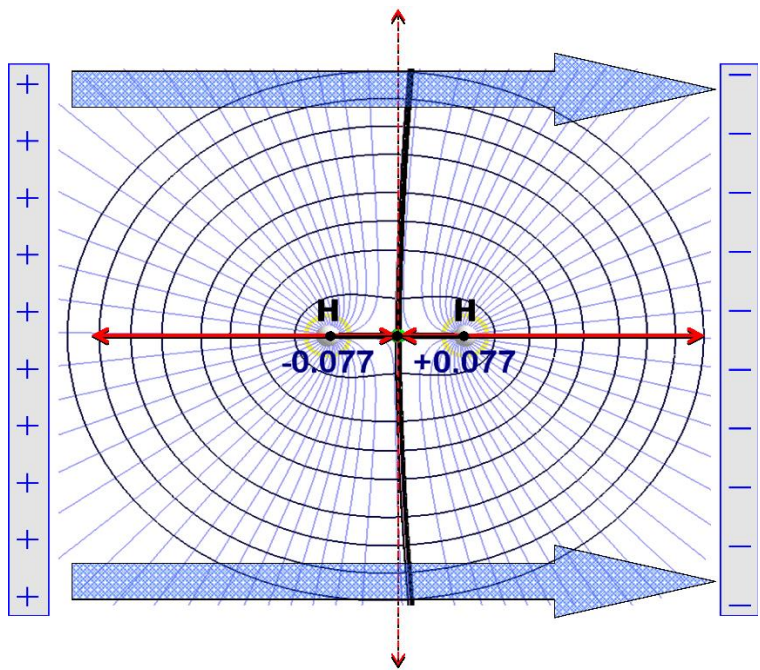

**Figure 3**
Electron density contour map and the associated gradient vector field of H$_2$ in the presence of an external electric field of $1.03 \times 10^{10}$ V.m$^{-1}$ (= 0.02 a.u.) pointing in the direction of the blue arrows. The interatomic surface intersecting the plane is represented by the thick line which is slightly concave toward the right. The values of the contours from outside inward are: 0.001, 0.002, 0.004, 0.008, 0.02, 0.04, 0.08, and 0.2 a.u. The numerical values are the field-induced atomic charges in a.u. The field-free optimized bond length $d$(H–H) = 0.7426 Å (not shown) and 0.7446 Å in the field of the figure. The vertical double-headed straight-line arrow passing through the bond critical point (BCP) contrasts with the field-induced curvature of the interatomic surface. The horizontal double-headed red arrows that meet at the BCP have *identical* lengths of 1.62 Å each. The horizontal double-headed red arrow to the right equals the length from the BCP to the $\rho$ = 0.001 au isodensity envelope but is unable to reach the same isodensity envelope on the left which it is 1.77 Å away from the BCP along the same line. The volume of the atom to the left enclosed by the $\rho$ = 0.001 au envelope is 65.54 a.u. (bohr$^3$) while the atom to the right has a volume of 54.70 a.u. The three black dots, two at the position of the nuclei and one at the BCP are separated by two thin black segments of identical lengths and which are each equal to the bonded radius of the atom to the positively-charged atom (the atom to the right) $r_b$ = 0.3633 Å, shorter than the bonded radius of the negative atom on the left $r_b$ = 0.3813 Å.

The H−H bond length is stretched in the field by about 0.0019 Å with an accompanying substantial and opposite changes in the two bonded radii: The bonded radius in the field-free H$_2$ molecule is half of the bond length, that is, 0.7446/2 = 0.3713 Å. The bonded radius of the hydrogen atom facing the negative pole (the positively charged H to the right of Figure 3) is compressed to 0.3633 Å and that of the negatively charged H atom extended to 0.3813 Å. As can be seen from Table 2, and with the exception of O$_2$, the changes in the bonded radii for the other homo-nuclear diatomic molecules are qualitatively similar, i.e., the bonded radius of the negatively charged atom lengthens while that of the positively charged atom shortens. For $O_2$, however, *both* bonded radii lengthen under the field, but more so for the negatively charged atom (Table 2). These observations mean that the interatomic surface is pulled closer to the nucleus of the positive atoms and further away from the negative one. In other words, *the zero-flux inter-atomic surface appears to be attracted by the negative pole of the external field, or said differently, is convex with respect to the direction of the external field*. This is understandable since the lower contours of the electron density are more polarizable and easier to pull by the positive end of the field and pushed by the field's negative end.

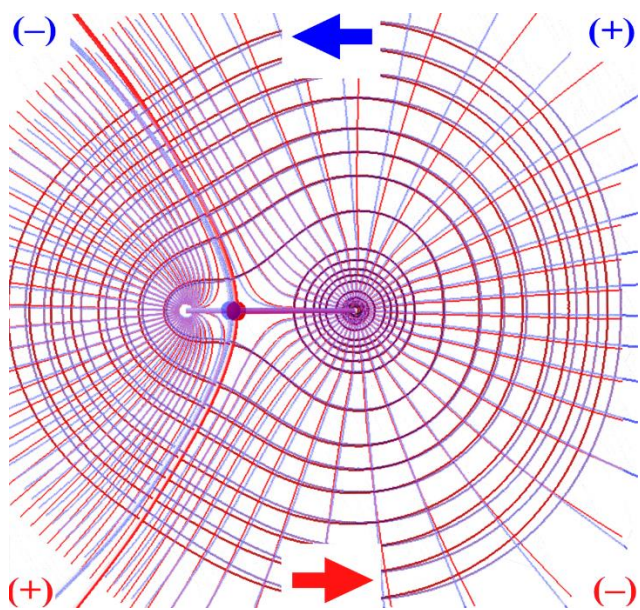

**Figure 4**
Superimposed electron density contour maps of HCl in the presence of $1.03 \times 10^{10}$ V.m$^{-1}$ (= 0.02 a.u.) external field in directions that are parallel (field oriented right-to-left, $\rho$ in blue contours) and antiparallel (field oriented left-to-right, $\rho$ in red contours) to the permanent dipole moment. The gradient vector field lines and interatomic surfaces have the same colors as their respective associated electron densities (the H basin is the one to the left). The superimposition has been accomplished by matching the inner contours of the Cl core. The values of the contours from outside inward are: 0.001, 0.002, 0.004, 0.008, 0.02, 0.04, 0.08, 0.2, 0.4, 0.8, 2.0, 4.0, 8.0 a.u. The optimized bond lengths [$d$(H–Cl)] in Å are 1.2736 for the field-free case (not shown), 1.2828 (parallel, blue), and 1.2702 (antiparallel, red).

No such a clear pattern emerges from the examination of hetero-nuclear diatomic molecules, except that the density toward the positive plate is more diffuse under the field than in the field-free case and, to the contrary, the density of the atom facing the negative plate is compacted. These effects can be seen by the superimposition of the electron density contour lines and their associated gradient vector field lines of the most polarizable hetero-diatomic, HCl, in the presence of two opposing fields of equal magnitudes (Figure 4).

External fields alter both the geometry (the set of nuclear positions $\mathbf{R}_i$) and the electron density, which are, of course, coupled through the Hellman-Feynman theorem.[53-54] The molecular electrostatic potential (MESP) in terms of the point-like nuclei and continuous electron density distribution can be written for a diatomic (in a.u.) as,[55]

$$V(\mathbf{r}) = \frac{Z_1}{|\mathbf{R}_1 - \mathbf{r}|} + \frac{Z_2}{|\mathbf{R}_2 - \mathbf{r}|} - \int \frac{\rho(\mathbf{r}')}{|\mathbf{r}' - \mathbf{r}|} d\mathbf{r}' + V_{\text{external}}(\mathbf{r}), \quad (2)$$

where whenever $\mathbf{R}_i = \mathbf{r}$, one eliminates the corresponding term of the first two. In equation (2) one adds any externally-imposed potential as the fourth term, in this case $V_{\text{external}}(\mathbf{r}) = V_{\text{external}}(z)$ since it is only a function of the z-coordinate in the chosen system orientation.

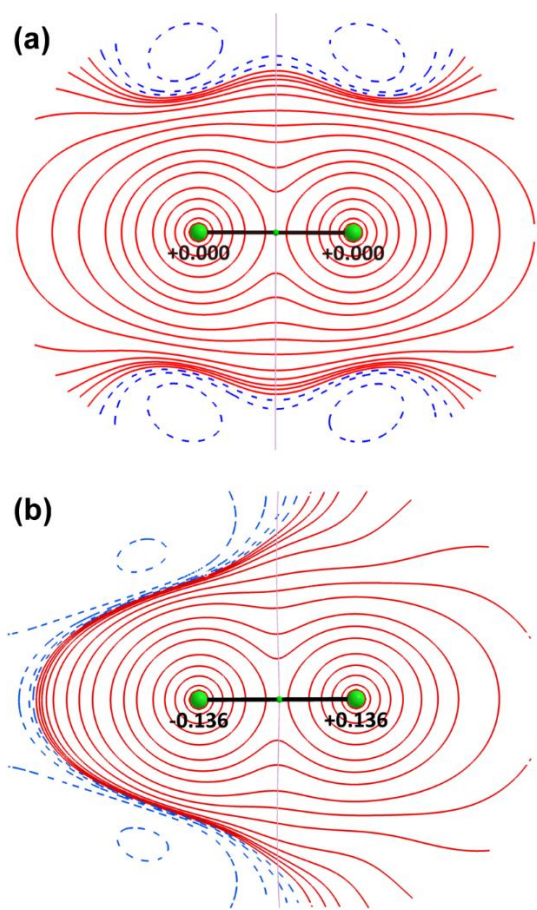

**Figure 5**
(a) The molecular electrostatic potential (MESP) of $Cl_2$ in absence of external fields and (b) the MESP of this molecule generated by the field-distorted total charge density (electronic and nuclear) but without the potential associated with the external field. In (b), the external field giving rise to the distorted density (not shown) is directed from left to right parallel to the inter-nuclear axis. The solid (red) contours are of positive values of the MESP and the dashed broken (blue) lines denote negative MESP contours. The values of these contours starting at the nodal surfaces are (in atomic units, a.u.): 0.001, 0.002, 0.004, 0.008, 0.02, 0.04, 0.08, 0.2, 0.4, 0.8, 2.0, 4.0, 8.0, 20.0, 40.0, 80.0, for the positive region of the MESP, and -0.001, -0.002, -0.004, -0.008, -0.02 for the negative region, (1 a.u. of electrostatic potential = $E_h/e$ = 27.211 V). The line connecting the nuclei is the bond path, and the little green dot between the nuclei is the BCP, while the intersection of the interatomic surface of local zero-flux in the gradient vector field is the pale line bisecting the bond path.

The external field of strength $1.03\times10^{10}$ V.m$^{-1}$ has a marked effect on the shape of the MESP. This effect is shown in Figure 5 with the highly polarizable homonuclear diatomic Cl$_2$ taken as an example. The figure compares the MESP of an unperturbed Cl$_2$ molecule and one subjected to a perturbing field directed from left to right. The field changes both the topography and the topology of the MESP, as can be seen from the figure.

First, the field-free cylindrical symmetry is distorted into a conical one in the field-perturbed MESP as can be seen by comparing the shapes of the contours of parts (a) and (b) of Figure 5. Second, the topology is altered: In Figure 5 (a) one can distinguish two tori of negative MESP circling each Cl nucleus and two collinear mirror image σ-holes[56] of positive electrostatic potential collinear with the inter-nuclear axis. The two tori of negative MESP each possess cylindrical symmetry and are mirror images of each other through a plane bisecting the bond perpendicularly. The field-distorted system of nuclear and electronic charges gives rise to an MESP that has a conical shape with only one torus of negative MESP surrounding the negatively charged chlorine atom. These distortions in the MESP alter the preferred paths of approach of this molecule and a charged reactant.

### 3.3 *Localization and Delocalization Indices under External Fields*

The QTAIM localization (LI or $\Lambda(\Omega)$) and delocalization (DI or $\delta(\Omega, \Omega')$) indices are important descriptors that capture the electronic structure of a molecule.[57-58] The localization index accounts for the number of localized electrons within a given atomic basin whereas the delocalization index provides the number of shared electrons between two atoms. The two indices are not independent since they are related by the expression:[41, 50, 57-59]

$$N(\Omega_i) = \Lambda(\Omega_i) + \frac{1}{2}\sum_{j\neq i}^{n}\delta(\Omega_i,\Omega_j) = \int_{\Omega_i} \rho(\mathbf{r})d\mathbf{r}, \tag{3}$$

where $N(\Omega_i)$ is the total average electron population of the $i^{\text{th}}$ atom $\Omega_i$ which can be obtained either via the first equality or, independently, by directly integrating the electron density over the volume of the atomic basin by the last equality. The atomic charges are then obtained from:

$$q(\Omega_i) = Z_{\Omega_i} - N(\Omega_i), \tag{4}$$

where $Z_{\Omega_i}$ is the atomic number.

Figure 6 displays the change in the delocalization index as a function of the external field. As Figure 6 shows, the changes in DI follow the general trends of the response of $\rho_b$ to the field (Figure 2) but neither with the same relative magnitude nor with the same ordering between different molecules. Since the range of changes in $\rho_b$ values is small, the empirical exponential dependence of the delocalization density on the electron density[60] observed with a larger range of variation is not seen here. Instead, all five studied homo-nuclear diatomic molecules exhibit a linear correlation of $\Delta\delta(\Omega, \Omega')$ and $\Delta\rho_b$ with $r^2 = 1.000$ ($H_2$, $N_2$, and $Cl_2$) and $r^2 = 0.999$ ($O_2$ and $F_2$). The dependence of $\Delta\delta(\Omega, \Omega')$ on $\Delta\rho_b$ for the hetero-nuclear diatomic molecules exhibit qualitatively similar general tends except for HCl. Thus, the squared linear regression coefficients $r^2$ of the $\Delta\delta(\Omega, \Omega')$ *vs.* $\Delta\rho_b$ correlation are 0.986 (HF), 0.985 (CO), and 0.995 (NO), but only 0.799 for HCl. For the latter, the change in DI is found to correlate linearly with the change in bond distance ($r^2 = 0.962$), and the anomalous dependence of $\Delta\rho_b$ on $\Delta$BL is also reflected in $\Delta\delta(\text{Cl}, \text{Cl}')$ for antiparallel fields above approximately $5\times10^9$ V.m$^{-1}$.

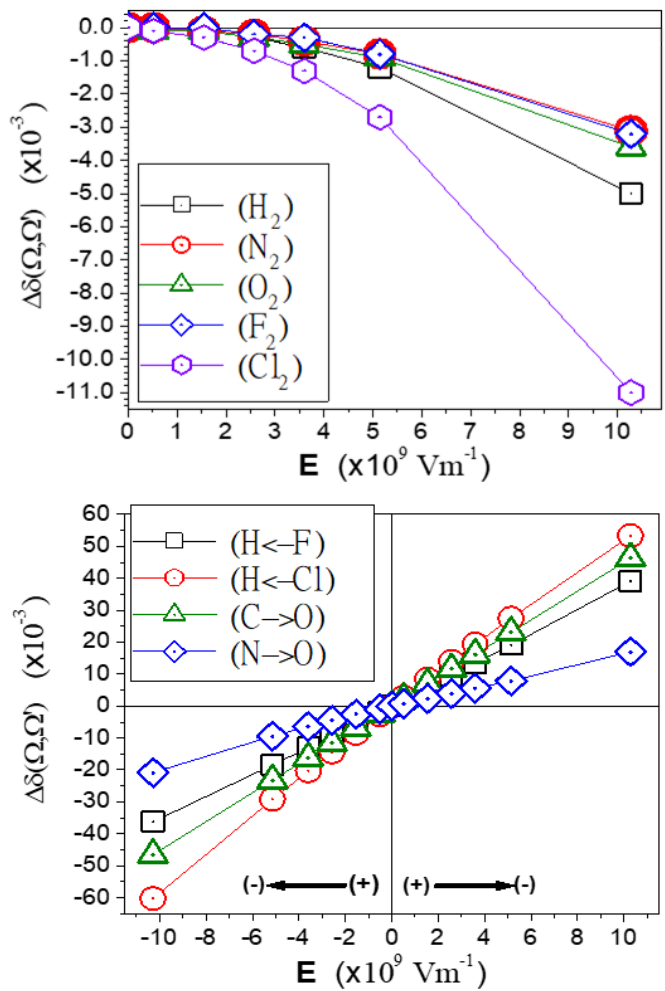

**Figure 6**
Plots of the change in the electron delocalization index (ΔDI), $\Delta\delta(\Omega, \Omega')$, at the optimized geometry as a function of the electric field (**E**) strength (in V.m$^{-1}$) for the homo-nuclear diatomics (*top*), and as a function of the field strength and direction for the hetero-nuclear diatomics (*bottom*). $\Delta\delta(\Omega, \Omega') = \delta(\Omega, \Omega')_E - \delta(\Omega, \Omega')_0$, where $\delta(\Omega, \Omega')_E$ is the DI in the field and $\Delta\delta(\Omega,\Omega')_0$ in the absence of external fields.

Since Δδ constantly changes in the same direction with the field for both homo- or hetero-diatomic molecules (Figure 6), we base the following discussion on the results with respect to the most extreme field intensity ($1.03\times10^{10}$ V.m$^{-1}$). For homo-diatomic molecules, $\Omega(2)$ (the atom near the positive plate) gains electronic charge and a more localized electron population is being attracted by the positive pole of the field in that basin and, consequently, this atom exhibits $\Delta\Lambda(\Omega(2)) > 0$. This rise in electron localization in the negatively charged atom comes at the expense of the localization of the positively charged atom which experiences a loss of electron localization under the effect of the external field, i.e., the localization index decreases for $\Omega(1)$ (or, $\Delta\Lambda(\Omega(1)) < 0$).

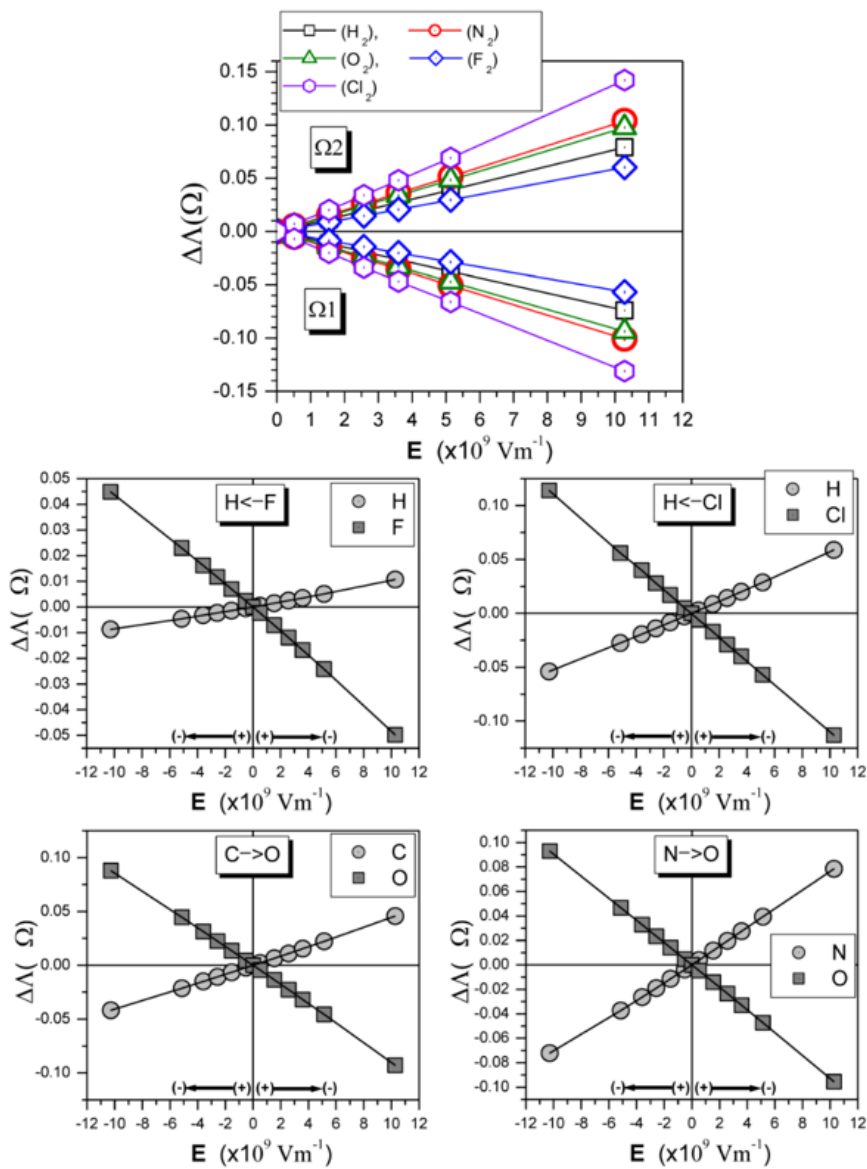

**Figure 7**
Plots of the change in the electron localization index ($\Delta$LI), $\Delta\Lambda(\Omega)$, at the optimized geometry as a function of the applied electric field (**E**) strength (in V.m$^{-1}$) for the homo-nuclear diatomics (*top*), and as a function of the field strength and direction for the hetero-nuclear diatomics (*middle* and *bottom*). $\Delta\Lambda(\Omega) = \Lambda(\Omega)_\mathbf{E} - \Lambda(\Omega)_0$, where $\Lambda(\Omega)_\mathbf{E}$ is the LI in the field and $\Lambda(\Omega)_0$ in its absence.

Interestingly, the increase in localization for Ω(2) is slightly larger than the decrease for Ω(1) leading to a net negative change in DI for a homo-diatomic molecule (Table 2 and Figure 7). But again one witnesses the operation of a type of field-induced compensatory response that largely cancels out at the molecular level. *The field-induced charge separation in homonuclear diatomics thus reduces electron sharing between the two atoms.* This can be described as decreasing the covalent nature of the bond accompanying the introduction of some polar-to-ionic character.

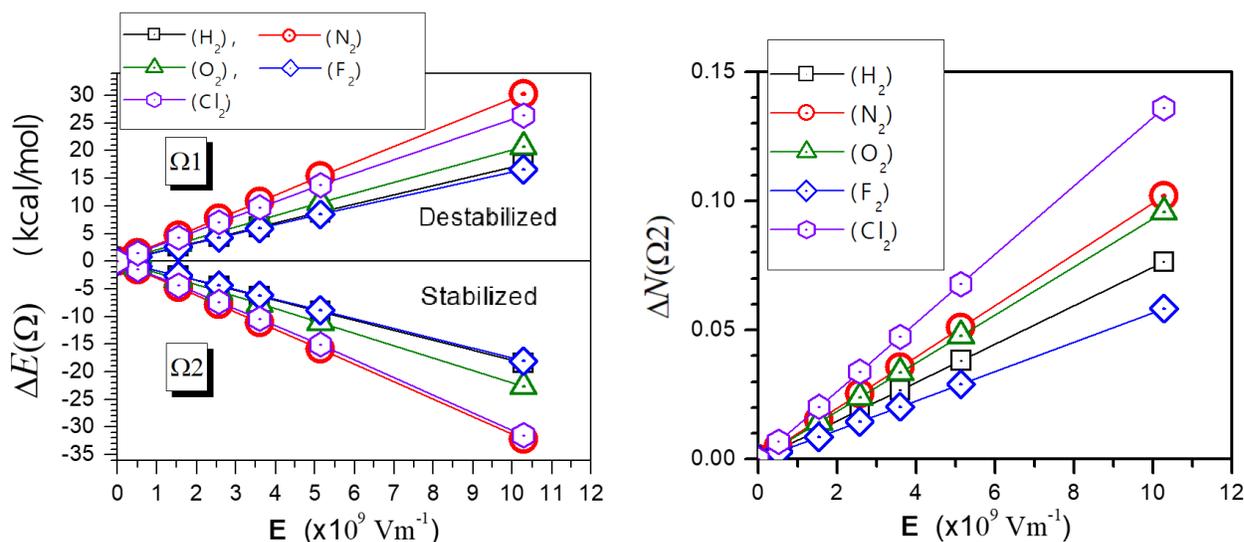

**Figure 8**
Plots of the change in the QTAIM atomic energies, $\Delta E(\Omega)$, as a function of the electric field (**E**) strength (in V.m$^{-1}$) (*left*) and the accompanying changes in the atomic electron populations $\Delta N(\Omega)$ (*right*) for the stabilized atoms (Ω2) of the homo-nuclear diatomics. $\Delta P(\Omega) = P(\Omega)_E - P(\Omega)_0$, where $P(\Omega)_E$ is the property (either $E$ or $N$) of atom Ω in the field and $P(\Omega)_0$ in the absence of external fields. Since atomic energies are always negative, a destabilized atom exhibits $\Delta E(\Omega) > 0$ and one that is more stable $\Delta E(\Omega) < 0$, but an atom that gains electron population (i.e. is more negatively charged/less positively charged) exhibits $\Delta N(\Omega) > 0$ and *vice versa*.

We use HF as an example for hetero-diatomic molecules, since all studied molecules exhibit qualitatively similar field-response with respect to electron localization and delocalization. The antiparallel electric field (right half of the plot, Figure 7) slowly increases electron localization on the hydrogen basin while it rapidly decreases localization for the fluorine basin. Under a parallel electric field (the left half of the plot in the left-middle panel of Figure 7), electrons become dramatically more localized in the fluorine basin and simultaneously less localized (but to a much slower degree) in the basin of the hydrogen atom. Similar to a homo-diatomic, the difference between the sum of atomic localization indices under the field and in the field-free case is reflected in the change observed in DI for the molecule (Figure 6). Thus, the H–F bond "becomes more covalent" with an increase in electron delocalization in an anti-parallel electric field and it becomes more polar in a parallel field.

As is seen in Figure 7, under a given external electric field strength, the induced difference in the localization indices of atomic basins are more pronounced for hetero-diatomics. Thus, molecules consisting of atoms with different localization indices (i.e., permanent bond dipole moments) demonstrate a greater response to external electric fields. The field-induced change in the atomic net charge and polarity of the molecule is of a great significance since it would not only affect the inter-molecular interactions, but can also be used to control chemical reactivity of a compound through strengthening or weakening the nucleophilic/electrophilic centers of the molecules.

### 3.4   *Atomic Charges and Atomic Virial Energies in External Fields*

Regardless of the molecule type, the changes in atomic charges and atomic energies closely relate to the changes observed in the localization index. Specifically, higher localization is correlated with larger atomic energy and larger absolute value of atomic charge. Thus, for a homo-diatomic applying an external electric field destabilizes the atom bearing an induced positive charge ($\Omega(1)$) (i.e. the energy of that atom is less negative that in the field-free case) as it loses charge density, while the field has opposite effects on the atom with an induced negative charge, that is, Ω(2) which is stabilized by the field (Figure 8).

For homo-diatomic molecules, the decrease (stabilizing effect) in the atomic energy of one atom is larger than the increase (destabilizing effect) for the other, and thus the total energy of the molecule decreases (gets stabilized) as has been previously discussed.[25, 61] Thus, an electric field applied on the $D_{\infty h}$ axis of a homo-diatomic molecule always stabilizes the molecule.

In the case of hetero-diatomic molecules (Figure 9), and following the changes in LIs and atomic energies, antiparallel fields destabilize HF and HCl and stabilize CO and NO. As was discussed above for HF, the decrease in electron localization on the fluorine basin as a result of the anti-parallel external electric field destabilizes the F atom, while the H basin is stabilized. Similar observations equally apply for HF and HCl as can be seen from the figure. In the case of CO and NO, the oxygen atom plays the role of the halogen and carbon or nitrogen the role of the hydrogen in the hydrogen halides.

Another atomic property related to the atomic energy is the atomic electron population. Figure 9 compares the change in the atomic energy of hetero-diatomics and the corresponding electron population of the atomic basins with the external electric field. As is seen, electron population in an atomic basin directly relates to the atomic energy, i.e., the larger the population, the more negative the energy the more stable is the atomic basin and a loss of population is accompanied with a decrease in stability, that is, less negative energy.

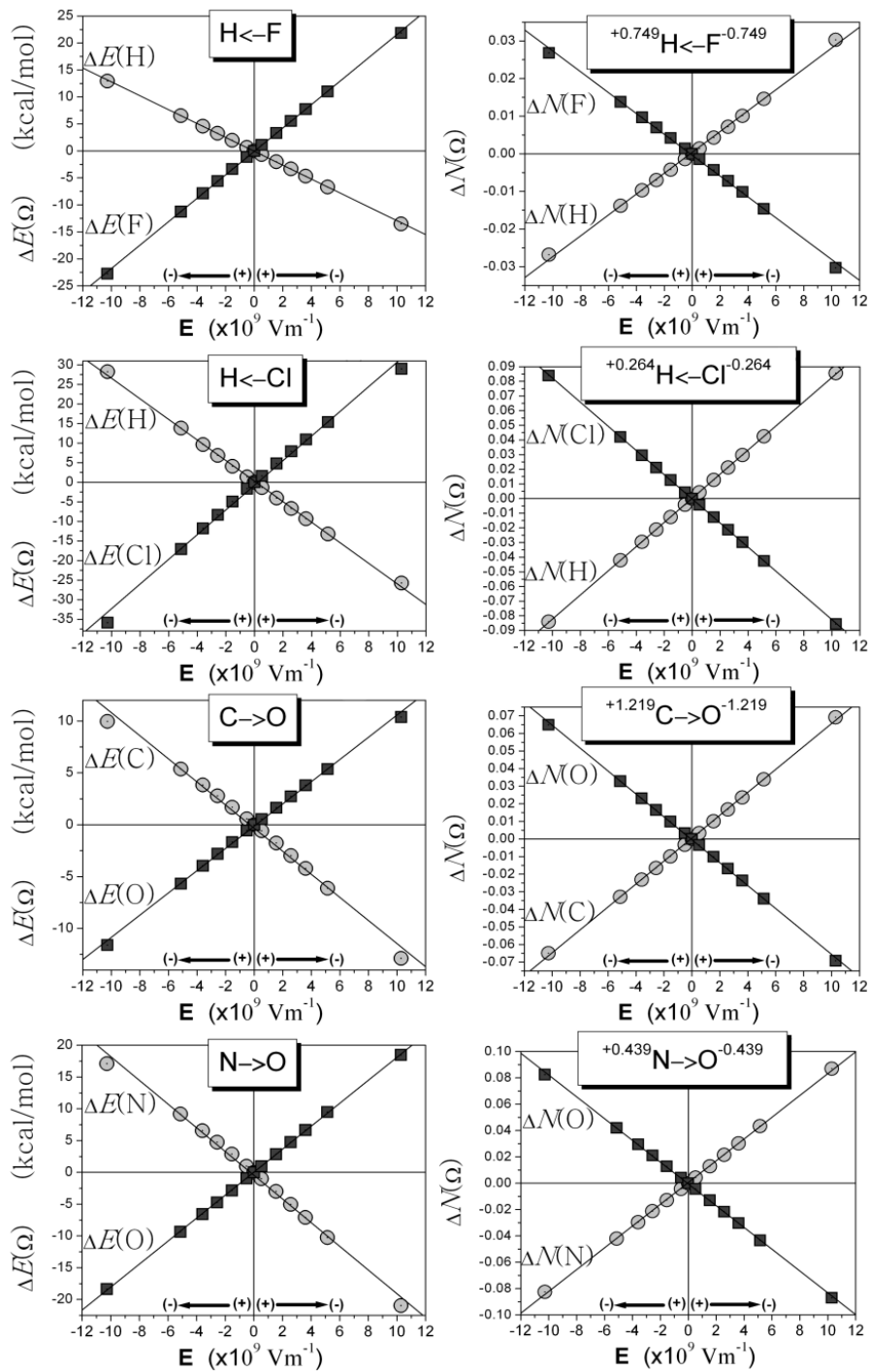

**Figure 9**
Plots of the change in the QTAIM atomic energies, $\Delta E(\Omega)$, as a function of the electric field (E) strength (in V.m$^{-1}$) (*left*) and the accompanying changes in the atomic electron populations $\Delta N(\Omega)$ (*right*) for the hetero-nuclear diatomics. $\Delta P(\Omega) = P(\Omega)_E - P(\Omega)_0$, where $P(\Omega)_E$ is the property (either *E* or *N*) of atom $\Omega$ in the field and $P(\Omega)_0$ in the absence of external fields. Since atomic energies are always negative, a destabilized atom exhibits $\Delta E(\Omega) > 0$ and one that is more stable $\Delta E(\Omega) < 0$, but an electron that gains electron population (i.e. is more negatively charged/less positively charged) exhibits $\Delta N(\Omega) > 0$ and *vice versa*. The inset at the top of each plot includes an arrow between the atomic symbols showing the direction of the permanent (field-free) dipole moment in the physicist convention while the inset at the top of each plot to the right is also labeled by the permanent QTAIM partial atomic changes in atomic units.

## 4. Conclusions

As the first step toward harnessing the physical and chemical properties of compounds and chemical processes by EEFs, the effects of external electric fields on atomic and bond properties of homo- and hetero-nuclear diatomic molecules have been elucidated. The results show that depending on the molecule and direction of the external field, electric fields can alter the ionic/covalent characteristics of the chemical bonds and thus the polarity of the molecule. The changes in the electron localization indices (gained by one basin and lost by the other) at a given electric field are not equal and the residual value adds to or subtracts from the delocalization index. This change manifests how an atomic property (localization index) alters a molecular (bond) property (i.e., delocalization index) under a given electric field.

The uniform electron distribution of homo-diatomic molecules is perturbed by external electric fields which leads to charge separation imparting a polar character and a reduced covalent nature of the bond. The change in the chemical bond character under external electric fields is characterized by the decrease in the electron density at BCP, a marked curvature of the zero-flux interatomic surface, a change in the atomic volumes, atomic energies, atomic populations, and descriptors of electron localization and delocalization. Several properties exhibit compensatory responses whereby the change in one atomic basin cancels the change in the other. These observations are more pronounced for the hetero-diatomic molecules, for which the electron distributions are uneven over the atomic basins. The above-mentioned changes can influence the nucleophilicity and electrophilicity of a molecule, hence providing means to manipulate and control certain chemical reactions. This assertion is based on previous work by Bandrauk *et al.* who showed that molecules exhibit sharp peaks near the transition state region in their dipole moments that arise primarily from the most polarizable atom or group in the system.[62] These peaks in the dipole moment surface along with the well-known peaks in polarizability near the transition-state region can be exploited to control the barriers of reaction through the proper choice of laser field frequencies, intensities, and phases.[63-64]

Electric fields are found to always stabilize homo-diatomic molecules (as anticipated from Eq. (1)). However, hetero-nuclear diatomic molecules may be stabilized or destabilized depending on the direction (parallel or anti-parallel) of the field. Atomic properties, i.e., atomic energy, atomic electron population, and localization index show a linear dependence on the external electric field strength, which make it possible to model and predict their behavior in a given field (within the studied range of linear behavior). On the other hand, correlations of molecular indices such as electron density at the bond critical point, bond length, and vibrational frequency (not discussed here, see Ref. [25]) with the fields are non-linear. The delocalization index is an exception and shows a linear correlation with the external field strength for homo-diatomics and a non-linear dependence for the hetero-diatomics.

The effects of electric fields on molecular and atomic properties arise from the redistribution of electrons and the distortion of the geometry (relative positions of the nuclei). This is reflected in the changes observed in the MESP, localization and delocalization indices, as well as in the atomic electron populations.

It is concluded that oriented external electric fields influence the charge distributions of atoms in molecules in a manner that can be understood with qualitative arguments. Molecules respond by attempting to minimize the change through compensatory field-induced effects so that the overall molecular effects, which are the sum of atomic-level effects, reflect opposing trends in the properties of the composing atoms.

**Conflicts of Interest Statement**

There are no conflicts of interest to declare.

**Acknowledgements**

Financial support of this work was provided by the Natural Sciences and Engineering Research Council of Canada (NSERC), NSERC Canada Research Chairs Program, Canada Foundation for Innovation (CFI), Western University, Mount Saint Vincent University, and Saint Mary's University.

Millam, N. J.; Klene, M.; Knox, J. E.; Cross, J. B.; Bakken, V.; Adamo, C.; Jaramillo, J.; Gomperts, R.; Stratmann, R. E.; Yazyev, O.; Austin, A. J.; Cammi, R.; Pomelli, C.; Ochterski, J. W.; Martin, R. L.; Morokuma, K.; Zakrzewski, V. G.; Voth, G. A.; Salvador, P.; Dannenberg, J. J.; Dapprich, S.; Daniels, A. D.; Farkas, Ö.; Foresman, J. B.; Ortiz, J. V.; Cioslowski, J.; Fox, D. J. Gaussian, Inc., Wallingford CT, 2009., Gaussian 09. **2010**.

47. Hovick, J. W.; Poler, J. C., Misconceptions in Sign Conventions: Flipping the Electric Dipole Moment. *Journal of Chemical Education* **2005,** *82* (6), 889.

48. CA, C., *Electricity*. Oliver and Boyd: London, 1961.

49. Keith, T. A., AIMAll. *http://aim.tkgristmill.com/* **2011**.

50. Bader, R. F. W.; Bayles, D., Properties of Atoms in Molecules: Group Additivity. *The Journal of Physical Chemistry A* **2000,** *104* (23), 5579-5589.

51. Cortés-Guzmán, F.; Bader, R. F. W., Role of functional groups in linear regression analysis of molecular properties. *Journal of Physical Organic Chemistry* **2004,** *17* (2), 95-99.

52. Cortés-Guzmán, F.; Bader, R. F. W., Transferability of group energies and satisfaction of the virial theorem. *Chemical Physics Letters* **2003,** *379* (1), 183-192.

53. Feynman, R. P., Forces in Molecules. *Physical Review* **1939,** *56* (4), 340-343.

54. Clusius, Einführung in die Quantenchemie. Von H. Hellmann. 350 S.,43 Abb., 35 Tab. Franz Deuticke, Leipzig u. Wien 1937. Pr. geh. RM. 20,-. geb. RM. 22. *Angewandte Chemie* **1941,** *54* (11-12), 156-156.

55. Bonaccorsi, R.; Scrocco, E.; Tomasi, J., Molecular SCF Calculations for the Ground State of Some Three-Membered Ring Molecules: (CH2)3, (CH2)2NH, (CH2)2NH2+, (CH2)2O, (CH2)2S, (CH)2CH2, and N2CH2. *The Journal of Chemical Physics* **1970,** *52* (10), 5270-5284.

56. Politzer, P.; Murray, J. S.; Clark, T., Halogen bonding and other σ-hole interactions: a perspective. *Physical Chemistry Chemical Physics* **2013,** *15* (27), 11178-11189.

57. Matta, C. F., Modeling biophysical and biological properties from the characteristics of the molecular electron density, electron localization and delocalization matrices, and the electrostatic potential. *Journal of Computational Chemistry* **2014,** *35* (16), 1165-1198.

58. Matta, C. F., Molecules as networks: A localization-delocalization matrices approach. *Computational and Theoretical Chemistry* **2018,** *1124*, 1-14.

59. Fradera, X.; Austen, M. A.; Bader, R. F. W., The Lewis Model and Beyond. *The Journal of Physical Chemistry A* **1999,** *103* (2), 304-314.

60. Matta, C. F.; Hernández-Trujillo, J., Bonding in Polycyclic Aromatic Hydrocarbons in Terms of the Electron Density and of Electron Delocalization. *The Journal of Physical Chemistry A* **2003,** *107* (38), 7496-7504.

61. Fouad, N. A., Analytical Insight into the Effect of Electric Field on Molecular Properties of Homonuclear Diatomic Molecules. *Current Smart Materials* **2017,** *2* (2), 153-161.

62. Matta, C. F.; Sowlati-Hashjin, S.; Bandrauk, A. D., Dipole Moment Surfaces of the CH4 + •X → CH3• + HX (X = F, Cl) Reactions from Atomic Dipole Moment Surfaces, and the Origins of the Sharp Extrema of the Dipole Moments near the Transition States. *The Journal of Physical Chemistry A* **2013,** *117* (32), 7468-7483.

63. Bandrauk, A. D.; Sedik el, W. S.; Matta, C. F., Effect of absolute laser phase on reaction paths in laser-induced chemical reactions. *J Chem Phys* **2004,** *121* (16), 7764-75.

64. Bandrauk, A. D.; Sedik, E. L. W. S.; Matta, C. F., Laser control of reaction paths in ion–molecule reactions. *Molecular Physics* **2006,** *104* (1), 95-102.
30